\documentclass[12pt]{iopart}

\usepackage{graphics}
\usepackage{epsfig}
\usepackage{iopams}
\usepackage{color}
\usepackage{pdfsync}
\usepackage[thinspace,amssymb]{SIunits}

\newcommand{\csb}{$\boldsymbol{B}\, \bot \, \boldsymbol{c}$\ }
\newcommand{\cpb}{$\boldsymbol{B}\, \| \, \boldsymbol{c}$\ }
\newcommand{\db}{$\Delta B$}

\pdfoutput=1

\begin{document}

\title[Anisotropic ESR of YbIr$_2$Si$_2$]{Anisotropic electron spin resonance of YbIr$_2$Si$_2$}
\author{T. Gruner, J. Wykhoff, J. Sichelschmidt, C. Krellner, C. Geibel, F. Steglich}
\address{Max Planck Institute for Chemical Physics of Solids, D-01187 Dresden, Germany}
\ead{Sichelschmidt@cpfs.mpg.de}

\date{\today}

\begin{abstract}
A series of electron spin resonance (ESR) experiments were performed on a single crystal of the heavy fermion metal YbIr$_2$Si$_2$\ to map out the anisotropy of the ESR-intensity $I_{\rm ESR}$ which is governed by the microwave field component of the $g$-factor. The temperature dependencies of $I_{\rm ESR}(T)$ and $g(T)$ were measured for different orientations and compared within the range \unit{2.6}{\kelvin} $\leq T \leq$ \unit{16}{\kelvin}. The analysis of the intensity dependence on the crystal orientation with respect to both the direction of the microwave field and the static magnetic field revealed remarkable features: The intensity variation with respect to the direction of the microwave field was found to be one order of magnitude smaller than expected from the $g$-factor anisotropy. Furthermore, we observed a weak basal plane anisotropy of the ESR parameters which we interpret to be an intrinsic sample property.
\end{abstract}

\maketitle
\section{Introduction}
The heavy fermion metal YbIr$_2$Si$_2$ \cite{hossain05a} is a rare example of a dense Kondo system that displays a well defined electron spin resonance (ESR) signal well below the Kondo temperature $T_K \cong$ \unit{40}{\kelvin} \cite{sichelschmidt07a}. This unexpected observation, discovered for the first time in YbRh$_2$Si$_2$ \cite{sichelschmidt03a}, initiated a series of theoretical approaches to understand the origin of the ESR line from both an picture of itinerant heavy electrons \cite{abrahams08a,wolfle09a,schlottmann09a,zvyagin09b} and a picture of localized Yb$^{3+}$ spins \cite{schlottmann09a,kochelaev09a}. The latter is supported by the localized character of the magnetic susceptibility \cite{kutuzov08a} and by the unambiguously local property of the observed ESR line, reflecting a large magnetic anisotropy influenced by the crystalline electric field  \cite{sichelschmidt07a,sichelschmidt07b}. Besides the relevance of this ESR line to study locally the physics of Kondo lattices it also provides the possibility to investigate the effect of a large $g$-factor anisotropy on the ESR intensity - a case which, although discussed thoroughly in terms of discriminating field-sweep and frequency-sweep ESR \cite{aasa75a,pilbrow84a}, to the best of our knowledge has never been reported for ESR probes in metals. Both, YbRh$_2$Si$_2$\ and YbIr$_2$Si$_2$ exhibit a highly anisotropic magnetic response, indicating that Yb$^{3+}$ moments are forming an easy plane square lattice within the crystallographic basal plane \cite{hossain05a,trovarelli00a}. Recently, the effect of the anisotropy in the Yb-Yb interactions has been considered for the resonant susceptibility within a molecular field model, yielding a satisfying description of the observed temperature dependence of the $g$-factor \cite{huber09a}. In this publication, we concentrate on a detailed investigation of the ESR-intensity of YbIr$_{2}$Si$_{2}$ for which, in contrast to YbRh$_{2}$Si$_{2}$, the resonance is experimentally observable for fields perpendicular to the basal plane. 

\section{Experimental Details}
\label{sec2}

\begin{figure}[htb]
  \centering
 \includegraphics[width=0.8\textwidth]{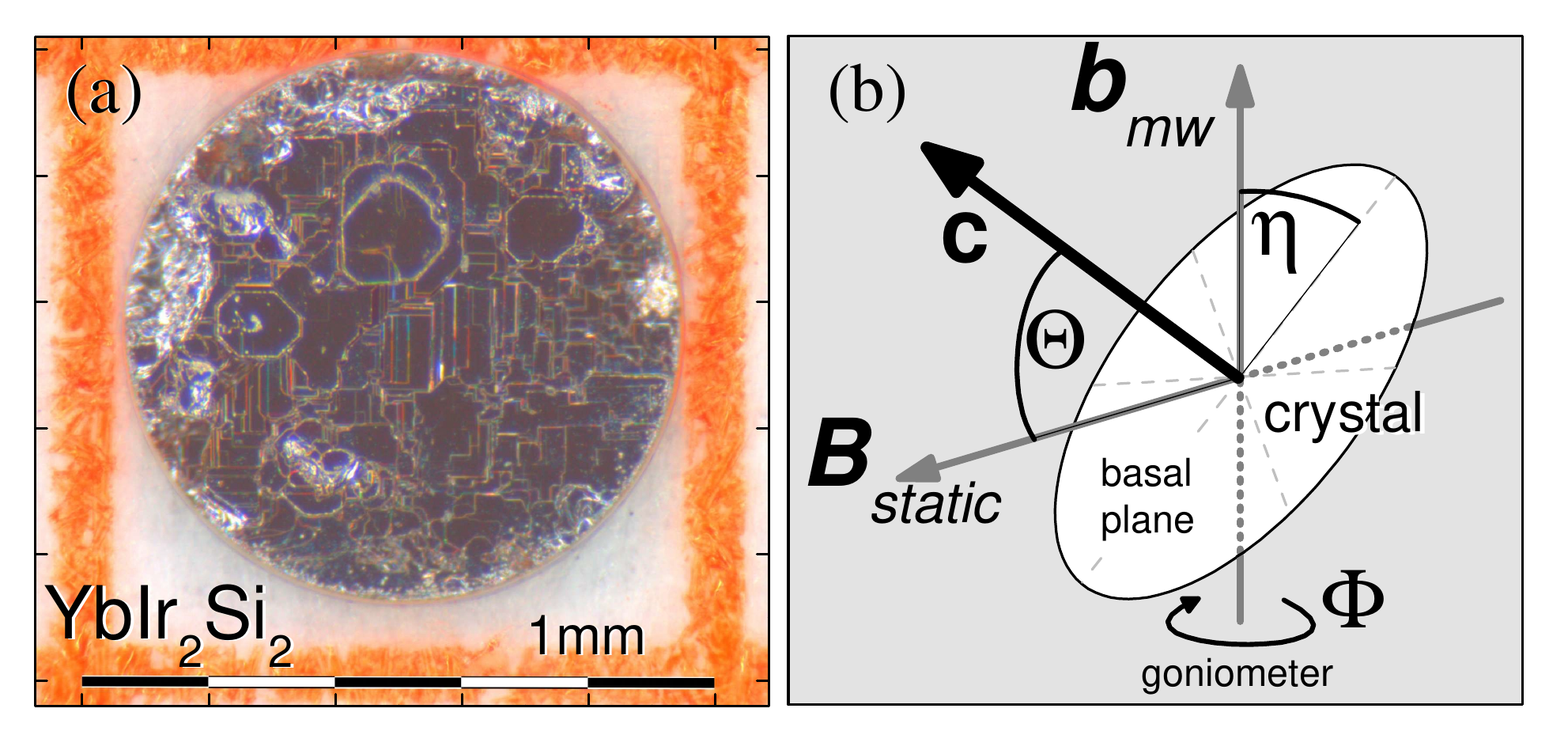}
  \caption{(a) Spark eroded single crystal YbIr$_2$Si$_2$\ with a circular disc diameter of \unit {890}{\micro \metre} used for the ESR investigations. (b) Sketch illustrates how the geometry of the measurements is defined by the angles $\Theta$, $\eta$ and $\Phi$ for a particular alignment of the sample crystallographic symmetry axis $\boldsymbol{c}$. The static magnetic field $\boldsymbol{B_{static}}\equiv \boldsymbol{B}$ is always perpendicular to the microwave magnetic field $\boldsymbol{b_{mw}}$.}
 \label{fig0}
\end{figure}

For our ESR measurements we used two In-flux grown single crystals of YbIr$_2$Si$_2$\, crystallizing in a body centered tetragonal structure (I-type) \cite{hossain05a}. A huge residual resistivity ratio ($\rho_{300K}/\rho_0$) of more than 200 indicates a high sample quality and very little crystalline disorder of the YbIr$_2$Si$_2$\ sample \cite{hossain05a}. The temperature dependence of the electrical resistivity and the magnetic properties have been thoroughly described elsewhere \cite{hossain05a,kutuzov08a}.\\ 
ESR detects the absorbed power $P$ of a magnetic microwave field $\boldsymbol{b_{mw}}$ as a function of a transverse external static magnetic field $\boldsymbol{B}$. To improve the signal-to-noise ratio, a lock-in technique is used by modulating the static field, which yields the derivative of the resonance signal $dP/dB$. All our ESR measurements were performed with a standard Bruker spectrometer at X-band frequencies ($\nu \approx$ \unit{9.4}{\giga \hertz}) and using a cylindrical resonator in TE$_{012}$ mode. The temperature was varied between \unit{2.6}{\kelvin} $\leq T \leq$ \unit{16}{\kelvin} using a He-flow cryostat.\\
We used a platelet-like single crystal where the crystallographic $\boldsymbol{c}$-axis was perpendicular to the platelet. The single crystal had a weight of $m\approx$ \unit{2.4}{\milli \gram}, a surface area of about \unit{1.4}{\milli \squaremetre} and a thickness up to \unit{150}{\micro \metre}. Figure ~\ref{fig0}(a) shows this crystal after cutting a circular disc by spark erosion. This reduces the mass to $m\approx$ \unit{0.6}{\milli \gram} for our measurements where we rotated the disc around its symmetry axis and keeping this axis aligned parallel to $\boldsymbol{b_{mw}}$. We expect geometry effects in the microwave absorption to be less important for this crystal shape. The spark erosion process did not affect the position, width, and shape of the ESR line.\\
%fig0
%%%%%%%%%%%%%%%%%%%%%%%%%%%%%%%%%%%%%%%%%%%%%%%%%%%%
%%%%%%%%%%%%%%%%%%%%%%%%%%%%%%%%%%%%%%%%%%%%%%%%%%%%
%
In Figure~\ref{fig0}(b) we show the three angles $\Theta$, $\eta$ and $\Phi$ introduced for describing the geometry of our measurement. To map the anisotropy we used on the one hand a goniometer to rotate the crystal by an angle $\Phi$ around the direction of the microwave field $\boldsymbol{b_{mw}}$. On the other hand we manually tilted the basal plane by an angle $\eta$ with respect to $\boldsymbol{b_{mw}}$ and affixed the such oriented YbIr$_2$Si$_2$\ crystal in paraffin.\\
For the further discussions we use the indices $\|$ and $\bot$ for the angles $0\degree$ and $90\degree$ between the crystallographic fourfold symmetry axis and the direction of the susceptibility component that determines the quantity in question. That means, the indices refer to the orientation of $\boldsymbol{b_{mw}}$ for the ESR-intensity $I_{\rm ESR}^{\bot, \|}$ whereas in the $g_{\bot, \|}$-factor they refer to the orientation of $\boldsymbol{B}$.
%fig1
%%%%%%%%%%%%%%%%%%%%%%%%%%%%%%%%%%%%%%%%%%%%%%%%%%%%
\begin{figure}[htb]
  \centering
 \includegraphics[width=0.8\textwidth]{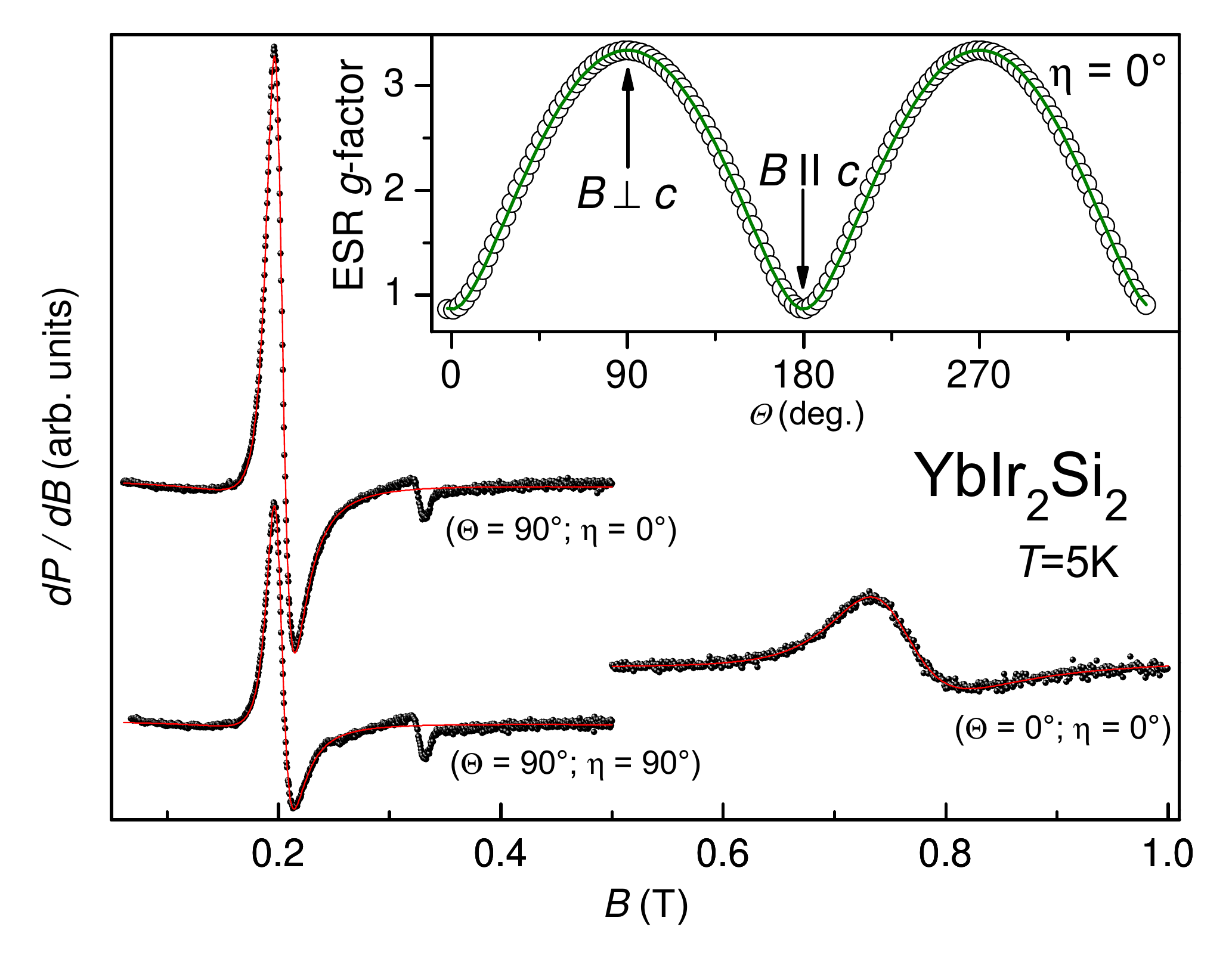}
  \caption{Typical ESR spectra of a YbIr$_2$Si$_2$\ single crystal at $T =$ \unit{5}{\kelvin} and for three different special the sample orientations. The small peak at $B \approx$ \unit{325}{\milli \tesla} arises from background effects of the microwave resonator. The lines represent fits to the data with a Dysonian line shape. In the inset the behavior of the ESR $g$-factor at $T =$\unit{5}{\kelvin} is shown. The sample was rotated around an axis lying in the basal plane parallel to $\boldsymbol{b_{mw}}$ (i.e. $\eta$ = \unit{0}{\degree}). The line fits the data with an uniaxial symmetry. Note that $\Theta = 0$ corresponds to \cpb \, and $\Theta$ = \unit{90}{\degree} to \csb.}
 \label{fig1}
\end{figure}
%%%%%%%%%%%%%%%%%%%%%%%%%%%%%%%%%%%%%%%%%%%%%%%%%%%%

\section{Experimental Results and Discussion}
\label{sec3}

Figure~\ref{fig1} presents examples of ESR signals of YbIr$_2$Si$_2$\ at \unit{5}{\kelvin} for three special crystal orientation. The line shape, resonance field and linewidth agree with the previously published spectra parameters \cite{sichelschmidt07a}. At $T =$ \unit{5}{\kelvin} the experimentally detected values are: $g_\bot(\unit{5}{\kelvin}) = 3.350 \pm 0.005$ and $g_\|(\unit{5}{\kelvin}) = 0.894 \pm 0.008$. The $\Theta$ dependence of the ESR $g$-factor can be nicely fitted with an uniaxial symmetry behavior $g(\Theta) = \sqrt{g_\|^2\cos^2\Theta + g_\bot^2\sin^2\Theta}$ (line in inset of figure~\ref{fig1}) reflecting the tetragonal YbIr$_2$Si$_2$\ crystal structure.  As one can see by the solid lines, all spectra nicely agree with a metallic Lorentzian, frequently called also Dysonian, because of its shape being identical to a conduction spin resonance in the limit of infinite spin diffusion time \cite{wykhoff07b}. From these fits we determined the ESR parameters linewidth \db, amplitude $amp$ and resonance field $B_{\rm res}$ (as given by the resonance condition $h\nu = g \mu_B \cdot B_{\rm res}$). The ratio $\alpha$, denoting the ratio of dispersion and absorption contributions in the Lorentzian shape, was kept constant, $\alpha = 0.97$, throughout the entire temperature range. A value $\alpha \approx 1$ is expected if the sample size is much larger than the microwave penetration depth. In our case the penetration depth of YbIr$_2$Si$_2$\ at $T =$ \unit{5}{\kelvin} is approximately \unit {0.8}{\micro \metre}.\\

\subsection{ESR-intensity and its temperature dependence}
\label{sec3.1}

The ESR-intensity $I_{\rm ESR}$ provides a measure for the static resonant susceptibility $\chi^R$, which is determined by the $g$-value component $g_1$ along the direction of the microwave magnetic field $\boldsymbol{b_{mw}}$ \cite{pilbrow84a}:

\begin{eqnarray}
\label{eq:0}
I_{\rm ESR} \propto \chi^R\propto g_1^2\;.
\end{eqnarray}

From the recorded ESR spectra $I_{\rm ESR}$ is determined by integrating the absorption curve in the frequency domain \cite{aasa75a}. The commonly used ESR setups record spectra by sweeping the magnetic field and leaving the microwave frequency constant. For this case the typical procedure to determine the intensity is also to integrate the field-sweep absorption curve which then yields a quantity $I_{\rm A}$ by calculating the area under the curve as \cite{wykhoff07b}:

\begin{eqnarray}
\label{eq:a}
I_{\rm A} \propto {\rm area} = \Delta B^2 \cdot amp \cdot \sqrt{1+\alpha^2}\;.
\end{eqnarray} 

However, for the proper relation between $I_{\rm A}$ and $I_{\rm ESR}$ it is necessary to be aware of the correct conversion between the ESR spectra integrals in the frequency- and field-domain. For systems with an effective spin $S=\frac{1}{2}$ \cite{aasa75a,pilbrow84a}
\begin{eqnarray}
\label{eq:b}
I_{\rm ESR}=I_{\rm A} \cdot g\;,
\end{eqnarray}
where $g$ corresponds to the $g$-value component along the static magnetic field $\boldsymbol{B}$ that also determines the resonance field via the resonance condition. Hence, especially for systems with large $g$-factor variation as a function of e.g. temperature or sample orientation the ESR intensity must be evaluated using equation (\ref{eq:b}) for a field-sweep experiment. For YbIr$_2$Si$_2$ one observes a very large $g$-factor anisotropy ($g_\bot \gg g_\|$) as shown in the inset of figure~\ref{fig1}. Thus, the dependence of the $g$-factor anisotropy on the intensity becomes important in an angle dependent experiment.\\

%Fig2
%%%%%%%%%%%%%%%%%%%%%%%%%%%%%%%%%%%%%%%%%%%%%%%%%%%%
\begin{figure}[htb]
  \centering
 \includegraphics[width=0.7\textwidth]{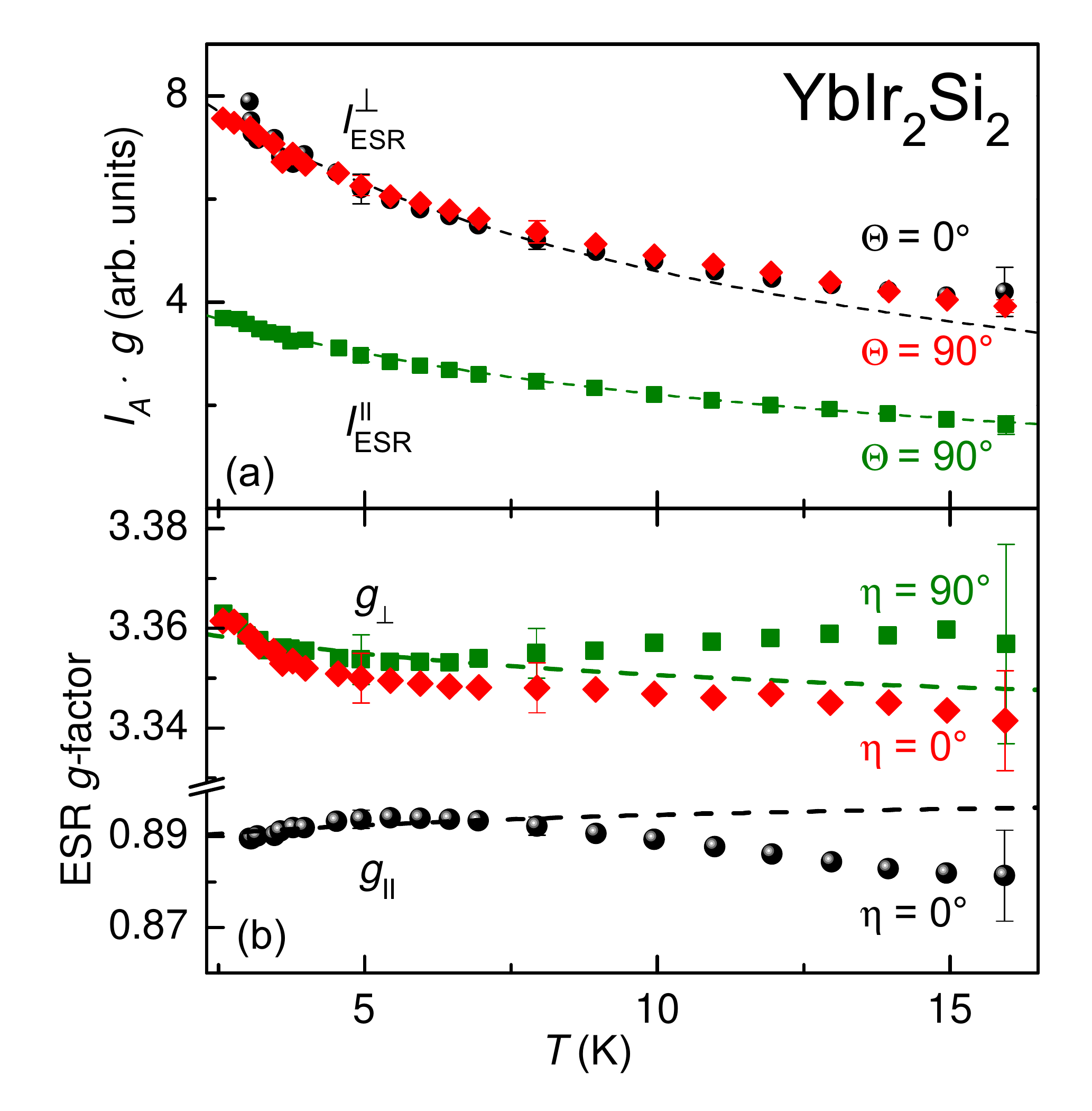}
  \caption{Temperature dependence for three special cases of (a) the ESR-intensity $I_{\rm ESR}=I_A \cdot g$ and (b) the corresponding $g$-factors. The $\|,\bot$ indices refer to angles as follows: $I_{\rm ESR}^{\|} \leftrightarrow \eta=\unit{90}{\degree}$\,, $I_{\rm ESR}^{\bot} \leftrightarrow \eta=\unit{0}{\degree}$ and $g_{\|}\leftrightarrow \Theta=\unit{0}{\degree}$\,, $g_{\bot}\leftrightarrow \Theta=\unit{90}{\degree}$. The dashed lines denote Curie-Weiss fits for the intensity and fits of the $g$-factor according equations~(\ref{eq:1}) and (\ref{eq:2}), see main text.}
 \label{fig2}
\end{figure}
%%%%%%%%%%%%%%%%%%%%%%%%%%%%%%%%%%%%%%%%%%%%%%%%%%%%
Figure~\ref{fig2}(a) shows the temperature dependence of $I_{\rm ESR}$, using the $g(T)$ data as shown in frame (b). Note that within the error bars $g_\bot$ does not depend on $\eta$ (figure~\ref{fig3} bottom panel). For a proper calculation of $I_{\rm ESR}(T)$ we took the microwave penetration depth $\propto 1/\sqrt{\sigma (T) \cdot \nu}$ ($\sigma$:electrical conductivity) into account. It is worth to note that according to equation~(\ref{eq:b}) the intensities
\begin{eqnarray*}
I^\perp_{\rm ESR}(\Theta = \unit{0}{\degree}; \eta =\unit{0}{\degree})&=&I_A \cdot g_\|(\Theta = \unit{0}{\degree}; \eta =\unit{0}{\degree}) \;\; {\rm and} \\
I^\perp_{\rm ESR}(\Theta = \unit{90}{\degree};\, \eta = \unit{0}{\degree})&=&I_A \cdot g_\perp(\Theta = \unit{90}{\degree};\, \eta = \unit{0}{\degree})
\end{eqnarray*}
are nearly equal, following the same temperature dependence and therefore behave as expected from equation (\ref{eq:0}) because $\boldsymbol{b_{mw}}$ is parallel to the basal plane in both configurations. The anisotropy of $I_{\rm ESR}$ in respect to the angle $\eta$ will be discussed in details in section \ref{intani}.\\

The dashed lines in figure~\ref{fig2}(a) represent Curie-Weiss laws $I_{\rm ESR}^{\bot, \|} \propto C_{\bot, \|}/(T+\theta_{\bot, \|})$ fitting the data between \unit{2.6}{\kelvin} $\leq T \leq$ \unit{6}{\kelvin}. The resulting Weiss temperatures are: $\theta_\bot \approx \unit{\left(8.6 \pm 0.6\right)}{\kelvin}$ and $\theta_\| \approx \unit{\left(8.8 \pm 0.4\right)}{\kelvin}$, i.e., within the experimental error of the intensity data one obtains $\theta_\bot \approx \theta_\|$. However, it is interesting to note that a small difference $\theta_\bot - \theta_\|$ may be the origin of the weak temperature dependence of $g(T)$ below 6K, see figure~\ref{fig2}(b). The molecular magnetic field description of the anisotropic Yb-Yb interaction by Huber \cite{huber09a,holanda09a} provides a link between the $g$-factor and the exchange anisotropy which is reflected in $\theta_\bot - \theta_\|$:
\begin{eqnarray}
\label{eq:1}
g_\bot(T) &=& g_{\bot}^0 \cdot \left(1-\frac{\theta_\bot - \theta_\|}{T + \theta_\bot}\right)^\frac{1}{2}\\
\label{eq:2}
g_\|(T) &=& g_{\|}^0 \cdot \left(1+\frac{\theta_\bot - \theta_\|}{T + \theta_\|}\right)
\end{eqnarray}
The dashed lines shown in figure~\ref{fig2}(b) represent fits of the $g$-factor temperature dependencies with equations~(\ref{eq:1}) and (\ref{eq:2}) for \unit{2.6}{\kelvin} $\leq T \leq$ \unit{6}{\kelvin}. We obtained for the adjustable parameters $g_{\bot}^0 \approx 3.34$ and $g_{\|}^0 \approx 0.90$. These values reasonably agree with the Yb$^{3+}$ $g^{\rm CEF}$-factors calculated for the $\Gamma_7$ Kramers doublet ground state in the crystal electric field (CEF) of YbIr$_2$Si$_2$\ : $g^{\rm CEF}_{\bot}=3.529$ and $g^{\rm CEF}_{\|}=0.929$ \cite{kutuzov08a}. The behavior of $g(T)$ for $T>6$~K is obviously inconsistent with the low temperature description according equations~(\ref{eq:1}) and (\ref{eq:2}). Instead, for instance, excited CEF levels may become relevant as observed in the linewidth temperature dependence \cite{sichelschmidt03a}.       

\subsection{Anisotropy of the ESR-intensity}
\label{intani}

We first explore the behavior of the ESR-intensity at $T=\unit{5}{\kelvin}$ by tilting the basal plane by an angle $\eta$ against $\boldsymbol{b_{mw}}$ (see figure~\ref{fig0}(b)). At the same time we carefully checked that the $\boldsymbol{c}$-axis is always perpendicular to the direction of the static magnetic field $\boldsymbol{B}$ (i.e. $\Theta = \unit{90}{\degree}$). 
%Fig3
%%%%%%%%%%%%%%%%%%%%%%%%%%%%%%%%%%%%%%%%%%%%%%%%%%%%
\begin{figure}[htb]
  \centering
 \includegraphics[width=0.8\textwidth]{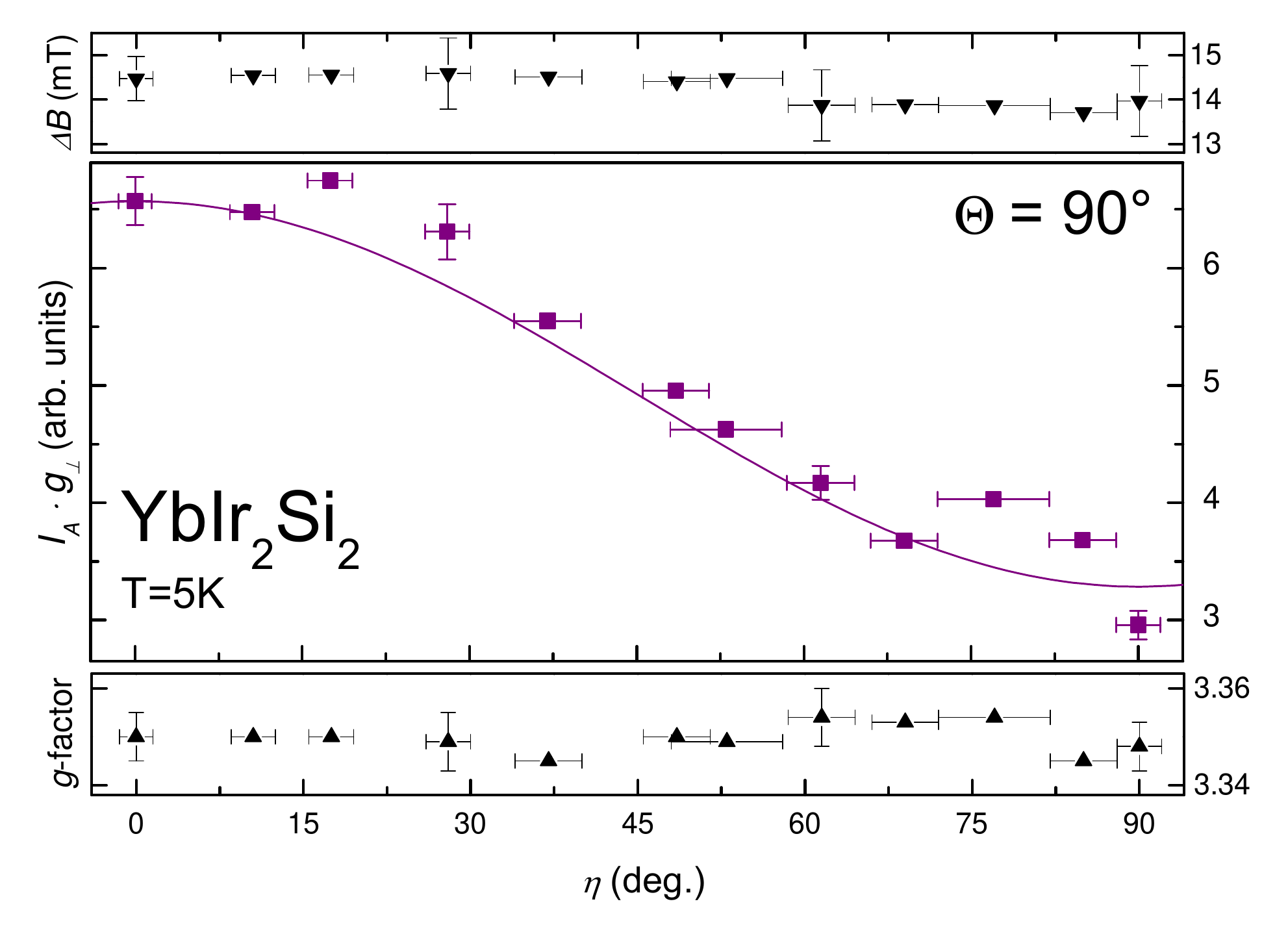}
  \caption{Variation of the intensity $I_{\rm ESR}$ by changing only the angle $\eta$ (see also figure \ref{fig0}b) and keeping $\boldsymbol{B}$ always perpendicular to $\boldsymbol{c}$-axis (i.e. $\Theta$=\unit{90}{\degree}). Solid line describes the data with an axial symmetry behavior, see main text. Upper and lower graph shows the corresponding linewidth \db\ and the $g_\bot$-factor, respectively.}
 \label{fig3}
\end{figure}
%%%%%%%%%%%%%%%%%%%%%%%%%%%%%%%%%%%%%%%%%%%%%%%%%%%%
As shown in the upper and lower frame of figure~\ref{fig3} the linewidth \db\ and the $g_\bot$-factor remain almost constant for this geometry of the measurement. This means that the amplitude of the absorption curve is the only parameter to determine $I_{\rm ESR}$ ($\alpha$ was kept constant for fitting the lineshape). The main frame of figure~\ref{fig3} shows the $\eta$ dependence of $I_{\rm ESR}$ with a periodicity of \unit{180}{\degree}. According to equation~(\ref{eq:0}) the ESR-intensity $I_{\rm ESR}$ is related to the $g$-value component $g_1$ along the microwave magnetic field $\boldsymbol{b_{mw}}$. Hence, $I^\bot_{\rm ESR}(\eta=0{\degree}) \propto g_\bot^2$ and $I^\|_{\rm ESR}(\eta=\unit{90}{\degree}) \propto g_\|^2$ and the axial symmetry angular dependence reads
$I_{\rm ESR}(\eta) = I(\eta=0{\degree})/g_\bot^2 \cdot \left[ g_\|^2\sin^2\eta + g_\bot^2\cos^2\eta\right]$. 
As shown by the solid line this equation describes the data with a ratio $I(\eta=0{\degree})/I(\eta=\unit{90}{\degree}) \approx 2$. However, from the $g$-values one would expect for the ratio $I^\bot_{\rm ESR}(\eta=0{\degree})/I^\|_{\rm ESR}(\eta=\unit{90}{\degree})=g_\bot^2 / g_\|^2 \approx 14$. This discrepancy cannot be justified by experimental errors and is also found for the system YbRh$_{2}$Si$_{2}$ where the difference is even much larger: $g_\bot^2 / g_\|^2 \approx 400$ as compared to the experimental value of $\approx 2.1$. At this point, we do not have any sound explanation for this discrepancy. 
%Fig4
%%%%%%%%%%%%%%%%%%%%%%%%%%%%%%%%%%%%%%%%%%%%%%%%%%%
\begin{figure}[htb]
  \centering
 \includegraphics[width=0.8\textwidth]{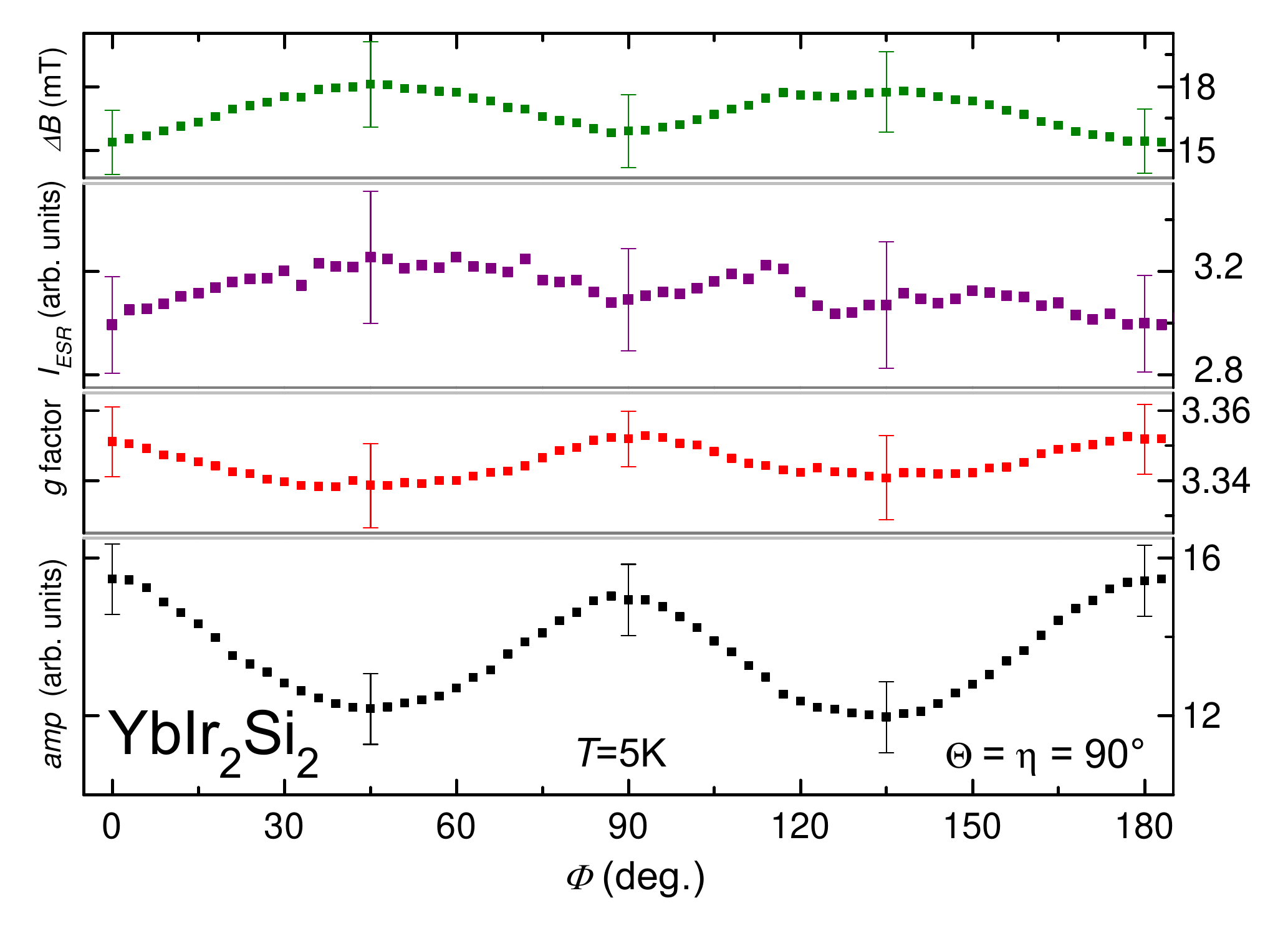}
  \caption{Variation of linewidth \db, intensity $I_{\rm ESR}$, $g$-factor and amplitude $amp$ by changing only the angle $\Phi$ (see also figure \ref{fig0}b) and always keeping the $\boldsymbol{c}$-axis  perpendicular to $\boldsymbol{B}$ and parallel to $\boldsymbol{b_{mw}}$ (i.e. $\Theta=\eta=$\unit{90}{\degree}). For this configuration a constant $g=g_\bot$ and a constant $I_{\rm ESR}=I^\|_{\rm ESR}\propto g_\|^2$ is expected.}
 \label{fig4}
\end{figure}
%%%%%%%%%%%%%%%%%%%%%%%%%%%%%%%%%%%%%%%%%%%%%%%%%%%

Finally we discuss the behavior of the ESR parameters by rotating the YbIr$_{2}$Si$_{2}$ crystal around its $\boldsymbol{c}$-axis by an angle $\Phi$, leaving the angles $\Theta = \eta = \unit{90}{\degree}$ constant. For this configuration we used the disc-shaped single crystal shown in figure~\ref{fig0}(a) where the $\boldsymbol{c}$-axis is pointing perpendicular to the disc plane. With this geometry the influence on the microwave field distribution should be minimized for the $\Phi$ rotation. We expect that the large $g$-factor anisotropy should neither be reflected in the resonance field nor in the intensity because the magnetic fields $\boldsymbol{B}$ and $\boldsymbol{b_{mw}}$ are always aligned perpendicular to the $\boldsymbol{c}$-axis or the crystallographic basal plane, respectively (compare figure \ref{fig0}(b)). Indeed, as shown in figure~\ref{fig4}, both $g$-factor (as determined by the resonance field) and $I_{\rm ESR}$ show only small variations which are consistent with a constant value within experimental accuracy. However, the \unit{90}{\degree}-periodicity of the anisotropy of $g$-factor, linewidth and the signal amplitude $amp$ is remarkable and is even exceeding the experimental error in the case of $amp$. This observation should not originate from sample geometry effects which are minimized by using a disc-shaped crystal. Also, we could confirm this anisotropy with a TE$_{102}$ rectangular resonator that has a different microwave resonant mode. Moreover, it is worth to note that the minima of the linewidth correlate with the $g$-factor maxima. Therefore, the anisotropic features point to a sample property although this effect is very weak and was not recognized in previous measurements: the $g$-value variation shown in figure~\ref{fig4} is three orders of magnitude smaller than in the usual case of rotation shown in figure~\ref{fig1}.

In order to explain the \unit{90}{\degree}-periodicity of the $g$-value one may assume a fourfold symmetry of a directional distribution of the $\boldsymbol{c}$-axes in respect to the rotation axis. This consideration holds only true if the spread of the $\boldsymbol{c}$-axes orientations of our crystal is $\approx \unit{4.7}{\degree}$ as estimated from the observed $g$-factor variation of $\approx 0.01$. However, the crystal mosaicity determined by a Laue-procedure shows a maximum deviation of $\unit{0.44}{\degree}$ from the expected $\boldsymbol{c}$-direction. Hence, only a weak intrinsic basal plane anisotropy of the $g$-value of YbIr$_2$Si$_2$ should be a plausible explanation of the \unit{90}{\degree}-$g$-periodicity. It indicates the influence of the off-diagonal crystalline electric field parameters, i.e. $B_{4}^{4}$ and $B_{6}^{4}$ which, for instance, are needed to describe $g$-factor anisotropies in cubic crystal symmetry \cite{abragam70a}.

\section{Summary}

We presented a detailed study of the anisotropy of the ESR signal on a YbIr$_2$Si$_2$\ crystal at a constant X-band frequency of \unit{9.4}{\giga \hertz}. As we used a standard field-sweep setup a proper determination of signal intensity $I_{\rm ESR}$ is required to take the large $g$-factor anisotropy into account. This leads to a basal plane intensity $I_{\rm ESR}^{\bot}$ which is independent from the angle $\Theta$ between crystalline c-axis and external magnetic field. Moreover, we found an almost perfect agreement in the temperature dependences of $I_{\rm ESR}^{\bot}(T)$ for $\Theta=\unit{0}{\degree}$ and $\Theta=\unit{90}{\degree}$.
The temperature dependence of the $g_{\|, \bot}$-factors anisotropies suggests a weak exchange anisotropy which is consistent with the behavior of $I_{\rm ESR}(T)$ within a molecular field treatment \cite{huber09a}. 
%This indicates that the peculiar temperature dependence of the $g$-factors is a consequence of the magnetic anisotropy itself and again points out that at least in the investigated temperature region the local properties of Yb$^{3+}$ moments determine the ESR signal. 
Furthermore, we obtained a remarkable result that is not expected for strongly anisotropic $g$-factors of the Yb$^{3+}$-ESR: a rotation of the basal plane by the angle $\eta$ respective the microwave field direction yields a much smaller change of $I_{\rm ESR}$ than expected from the $g$-factor anisotropy. Another interesting behavior was found when leaving the crystalline symmetry axis oriented with fixed angles $\Theta = \eta = \unit{90}{\degree}$ respective to the external field: then the rotation of the crystalline sample still showed a \unit{90}{\degree}-periodicity in the ESR line parameters indicating the influence of the off-diagonal crystalline electric field parameters.

\section*{Acknowledgements}

We thank Dave Huber and Elihu Abrahams for fruitful discussions which stimulated large parts of our investigations. We acknowledge the Volkswagen foundation (I/84689) for financial support.\\

%\section*{References}
%\bibliography{JoergBib}

\begin{thebibliography}{10}

\bibitem{hossain05a}
Hossain Z, Geibel C, Weickert F, Radu T, Tokiwa Y, Jeevan H, Gegenwart P and
  Steglich F 2005 {\em Phys.\ Rev.\ B\/} {\bf 72} 094411

\bibitem{sichelschmidt07a}
Sichelschmidt J, Wykhoff J, {H-A Krug von Nidda}, Fazlishanov I~I, Hossain Z,
  Krellner C, Geibel C and Steglich F 2007 {\em J. Phys. Cond. Mat.\/} {\bf 19}
  016211

\bibitem{sichelschmidt03a}
Sichelschmidt J, Ivanshin V, Ferstl J, Geibel C and Steglich F 2003 {\em Phys.\
  Rev.\ Lett.\/} {\bf 91} 156401

\bibitem{abrahams08a}
Abrahams E and W\"{o}lfle P 2008 {\em Phys. Rev. B\/} {\bf 78} 104423

\bibitem{wolfle09a}
W\"olfle P and Abrahams E 2009 {\em Phys. Rev. B\/} {\bf 80} 235112 (pages~8)

\bibitem{schlottmann09a}
Schlottmann P 2009 {\em Phys. Rev. B\/} {\bf 79} 045104

\bibitem{zvyagin09b}
Zvyagin A~A, Kataev V and B\"uchner B 2009 {\em Phys. Rev. B\/} {\bf 80} 024412

\bibitem{kochelaev09a}
Kochelaev B~I, Belov S~I, Skvortsova A~M, Kutuzov A~S, Sichelschmidt J, Wykhoff
  J, Geibel C and Steglich F 2009 {\em Eur. Phys. J. B\/} {\bf 72} 485--489

\bibitem{kutuzov08a}
Kutuzov A, Skvortsova A, Belov S, Sichelschmidt J, Wykhoff J, Eremin I,
  Krellner C, Geibel C and Kochelaev B 2008 {\em J. Phys. Cond. Mat.\/} {\bf
  20} 455208

\bibitem{sichelschmidt07b}
Sichelschmidt J, Wykhoff J, von Nidda H~A~K, Ferstl J, Geibel C and Steglich F
  2007 {\em J. Phys. Cond. Mat.\/} {\bf 19} 116204

\bibitem{aasa75a}
Aasa R and V\"anng\aa rd T 1975 {\em J. Magn. Res.\/} {\bf 19} 308

\bibitem{pilbrow84a}
Pilbrow J~R 1984 {\em J. Magn. Res.\/} {\bf 58} 186

\bibitem{trovarelli00a}
Trovarelli O, Geibel C, Mederle S, Langhammer C, Grosche F~M, Gegenwart P, Lang
  M, Sparn G and Steglich F 2000 {\em Phys.\ Rev.\ Lett.\/} {\bf 85} 626

\bibitem{huber09a}
Huber D~L 2009 {\em J. Phys. Cond. Mat.\/} {\bf 21} 322203

\bibitem{wykhoff07b}
Wykhoff J, Sichelschmidt J, Lapertot G, Knebel G, Flouquet J, Fazlishanov I~I,
  Krug~von Nidda H~A, Krellner C, Geibel C and Steglich F 2007 {\em Science
  Techn. Adv. Mat.\/} {\bf 8} 389

\bibitem{holanda09a}
Holanda L, Duque J, Bittar E, Adriano C, Pagliuso P, Rettori C, Hu R, Petrovic
  C, Maquilon S, Fisk Z, Huber D and Oseroff S 2009 {\em Physica B\/} {\bf 404}
  2964 -- 2968

\bibitem{abragam70a}
Abragam A and Bleaney B 1970 {\em Electron Paramagnetic Resonance of Transition
  Ions\/} (Oxford: Clarendon Press)

\end{thebibliography}
%\bibliographystyle{iopart-num}

\end{document}